# Guided Modes of Anisotropic van der Waals Materials Investigated by Near-Field Scanning Optical Microscopy

Daniel Wintz[1‡], Kundan Chaudhary[1‡], Ke Wang[2], Luis A. Jauregui[2], Antonio Ambrosio[3], Michele Tamagnone[1], Alexander Y. Zhu[1], Robert C. Devlin[1], Jesse D. Crossno[1], Kateryna Pistunova[2], Kenji Watanabe[4], Takashi Taniguchi[4], Philip Kim[2], Federico Capasso[1]

*[1]Harvard John A. Paulson School of Engineering and Applied Sciences, Harvard University, Cambridge, MA 02138, USA*

*[2]Department of Physics, Harvard University, Cambridge, MA 02138, USA*

*[3]Center for Nanoscale Systems, Harvard University, Cambridge, MA 02138, USA*

*[4]National Institute for Materials Science, Namiki 1-1, Tsukuba, Ibaraki 305-0044, Japan*

‡ These authors contributed equally to this work

Corresponding author: capasso@seas.harvard.edu

**Abstract:**

Guided modes in anisotropic two-dimensional van der Waals materials are experimentally investigated and their refractive indices in visible wavelengths are extracted. Our method involves near-field scanning optical microscopy of waveguide (transverse electric) and surface plasmon polariton (transverse magnetic) modes in h-BN/SiO$_2$/Si and Ag/h-BN stacks, respectively. We determine the dispersion of these modes and use this relationship to extract anisotropic refractive indices of h-BN flakes. In the wavelength interval 550-700 nm, the in-plane and out-of-plane refractive indices are in the range 1.98-2.12 and 1.45-2.12, respectively. Our approach of using near-field scanning optical microscopy allows for direct study of interaction between light and two-dimensional van der Waals materials and heterostructures.

Two-dimensional (2D) materials have recently garnered significant interest due to their high electrical mobility, atomic-level flatness, and large exciton binding energies which enable versatile nanophotonics and optoelectronics applications [1-23]. However, the understanding of the interaction of visible light surface plasmon polaritons with 2D materials and more complex van der Waals (vdW) heterostructures [5-7] is still in its infancy [9, 23]. In this work, we report a near-field scanning optical microscopy (NSOM) [25-28] study of highly crystalline hexagonal boron nitride (h-BN) [21] mechanically-exfoliated flakes at visible wavelengths, demonstrating the direct observation of (i) transverse electric (TE) waveguide modes supported by h-BN on $SiO_2$ and (ii) the interaction of transverse magnetic (TM) surface plasmon polariton (SPP) modes supported by silver (Ag) with a thin h-BN flake. From NSOM scans, we estimate the intrinsically anisotropic optical dielectric constants ($n_x = n_y \neq n_z$) of h-BN, which is an arduous task by conventional methods such as ellipsometry [24]. Our technique can be extended to other vdW solids and heterostructures, where we anticipate the study of guided modes coupled to 2D materials to be a useful tool in exploring rich physics of surface polaritons [9-14], plexcitons (SPPs-excitons) [29], gate-tunable [17, 30, 31], layer number dependent optical properties [13], in-plane anisotropy [15, 23], and selective circular dichroism [18, 19].

For our experiments, we fabricate waveguiding h-BN structures patterned using electron beam lithography (EBL, see Methods) on two different substrates. In the first case (Figure 1a), a 65 nm thin flake of h-BN is placed on top of a silicon (Si) substrate with 285 nm of thermal oxide ($SiO_2$). Because the refractive index of bulk h-BN is larger than the one of $SiO_2$, asymmetric TE dielectric slab waveguide modes supported by the h-BN are to be expected.

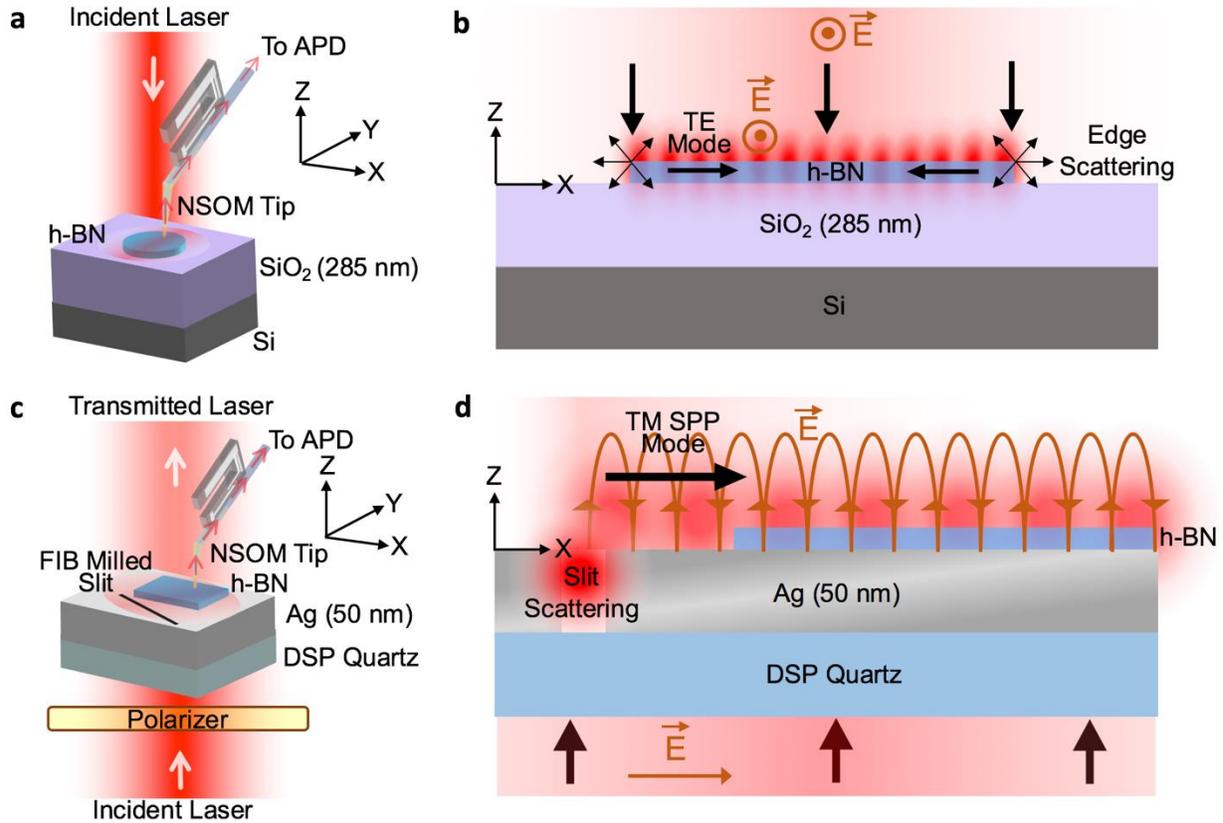

**Figure 1 | NSOM observation of guided modes in h-BN**. (**a**) TE modes observation with NSOM setup in collection mode and top-side laser illumination. The sample is a 65 nm thick h-BN flake on Si + 285 nm $SiO_2$ substrate. (**b**) Formation of fringes due to the interference of TE guided mode and the illuminating beam. The edges of the sample scatter the incident beam which subsequently launches the TE waveguide modes. The fringes are not due to the stationary waves in the flake, but rather due to the interference of the guided mode with the Gaussian beam focused on the sample. See Supplementary Information for more details. (**c**) TM SPP modes observation with NSOM setup in collection mode and back-side laser illumination. The sample is a 10 nm thick h-BN flake on 50 nm Ag on a double

side polished (DSP) quartz substrate. (**d**) Formation of fringes due to interference of TM SPP guided mode and the beam transmitted through the Ag layer. SPPs are launched by a slit created in the Ag layer by focused ion beam (FIB) milling, and subsequently propagate in the region of the sample covered with h-BN. In both geometries (**a**, **b**) and (**c**, **d**), the impinging and guided modes have the same polarization (electric field). Subsequently, the fringes arise from interference of excited guided modes and incident beam, and are directly observed by the NSOM probe as a variation of the local optical intensity.

Any guided mode is by definition evanescent, which means it cannot be observed with common optical microscopes that detect only far fields. To probe the near-fields on the samples, we use an NSOM in collection mode (Nanonics Imaging Ltd.), which utilizes a scanning metallized tapered fiber optic tip with a subwavelength aperture (effective diameter ~50 nm) [25-28]. An avalanche photodetector (APD) is connected at the other side of the fiber, such that the measured signal is proportional to the local optical intensity at the tip. A supercontinuum laser (NKT Photonics supercontinuum laser) is used to illuminate the sample from above with a filter (NKT Select) selecting a single wavelength with a 10 nm bandwidth in the spectral range from 550-700 nm. The impinging Gaussian beam, which can be approximated locally as a plane wave, can excite guided modes in the h-BN structure due to the scattering at the sample edges (Figure 1b). In the near-field region of h-BN, the optical electric field is the superposition of the impinging beam and the guided TE mode, creating interference fringes in the near-field which can be imaged by the NSOM tip [32, 33].

For the second case (Figure 1c), the flakes are placed on a 50 nm thin Ag film on DSP quartz, and SPPs are launched using a 200 nm wide slit (milled with FIB) illuminated from the back-side, as described in our previous work [32]. Similar to the first case, the NSOM can image the SPPs and fringes are visible due to the interference of the SPPs and the transmitted beam (Figure 1d). The h-BN flake is placed in the path of the SPP, and the resulting SPP mode, modified by the presence of the h-BN, can be studied with NSOM scans of the near-field of the sample.

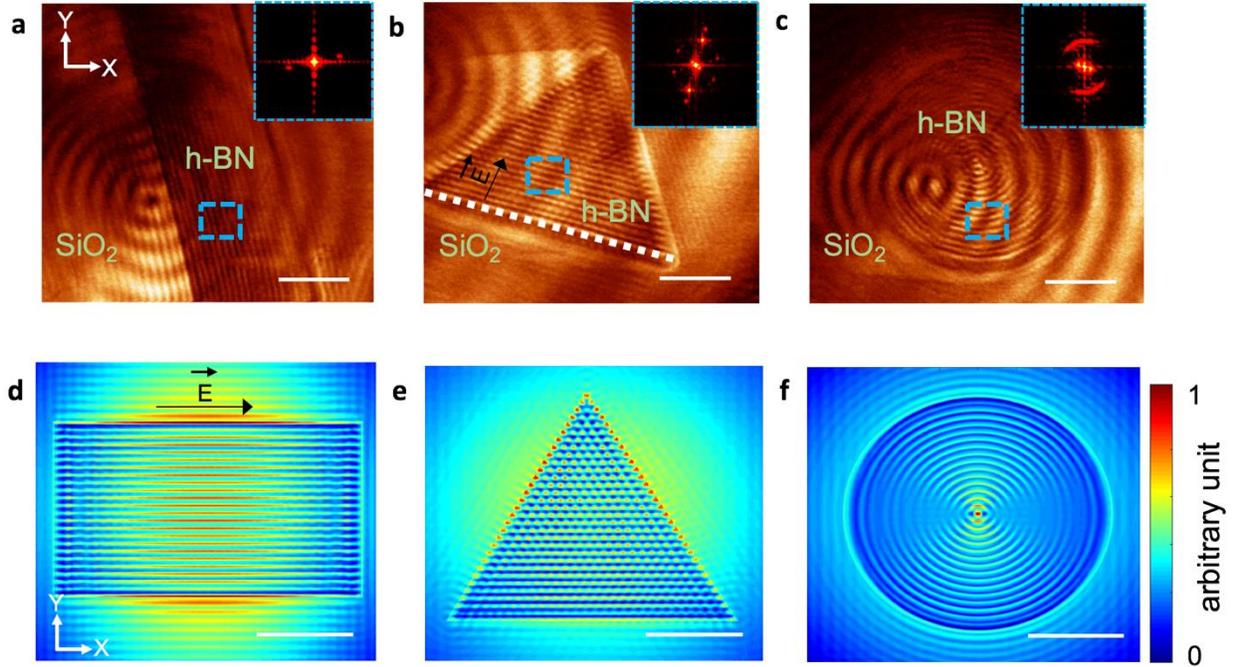

**Figure 2 | NSOM observation of TE waveguide modes in h-BN**. (**a-c**) Measurement of the optical intensity of interference fringes for $\lambda = 650\ nm$ in patterned flakes (rectangle, triangle, and circle, respectively). The larger fringes with greater inter-fringe spacing are the result of the focusing of the Gaussian beam (see Supplementary Information). Short period fringes (inside the h-BN region) correspond to high wavevector guided modes. Insets show the fast Fourier transform (FFT) of the corresponding region in the blue dashed rectangle. (**d-f**) Finite difference time domain (FDTD) simulation of the selected h-BN sample geometries, highlighting the guided wave pattern, with fringes spaced by $\lambda = \lambda_0/n_{eff}$ where $\lambda_0$ is the free space wavelength and $n_{eff}$ is the effective index of the waveguide mode. The excitation beam is polarized along x-axis. The hot spots on the triangle in (**b, e**) are due to the intersections of fringes originating from all the edges (see Supplementary Information for mathematical description). The fringes parallel to the edge in (b), highlighted by a dashed white line, are dominant with respect to fringes parallel to other edges. This is due to the fact that the incident beam is polarized and therefore the intensity of the fringes depends on their orientation with respect to the polarization of the incident beam. In particular, as described in Supplementary

Information, the intensity of each guided mode and its fringes is maximized when the incident polarization is parallel to the edge. Scale bars in (**a-f**) are 5 μm.

Figure 2 illustrates the experimental results for the first case, TE waveguide modes in a 65 nm thin h-BN flake on SiO$_2$. Because the mode is launched from the edges of the flake, the h-BN has been patterned in regular shapes to obtain uniform, easier to study interference patterns (Figure 2a-c). The NSOM scans reveal large circular fringes (associated with the focusing of the Gaussian beam on the sample) and, most importantly, a regular set of fringes on the h-BN, indicating the presence of waveguide modes.

The waveguide mode and the incoming beam interfere to form such fringes, which are parallel to the edges confirming their origin from edge scattering (see Supplementary Information). Non-guided scattered fields are also observed outside the sample as faint fringes immediately near the edges. Importantly, these fringes are not standing modes on the h-BN flakes. Rather, they originate from the interference of the incident beam with the guided modes launched by each edge and propagating away from it (see Supplementary Information for a mathematical description). To confirm the exact nature of these modes, FDTD simulations were performed on similar structures (Figure 2d-f). Modeling of h-BN flakes requires careful consideration of the optical anisotropy of the sample, as typical for layered materials, they show different refractive indices for the in-plane ($n_x = n_y$) and out-of-plane ($n_z$) directions [24]. Using the initial ansatz of $n_x = n_y = n_z = 2.0$ from bulk h-BN, similar fringes are found in the FDTD computed near-field (Figure 2d-f). The FDTD also shows that the guided mode is TE (see Supplementary Figure 2), and its presence can be explained by the fact that both air (above) and SiO$_2$ (below) have smaller refractive indices than h-BN. These conditions lead to the presence of confined TE dielectric waveguide modes. The modes in similar dielectric stacks are well known, and their TE nature is explained by the fact that TM modes are less confined and in the cutoff region for the h-BN thickness used here [34]. The same is true for higher-order TE modes, resulting in the propagation of first order TE modes alone.

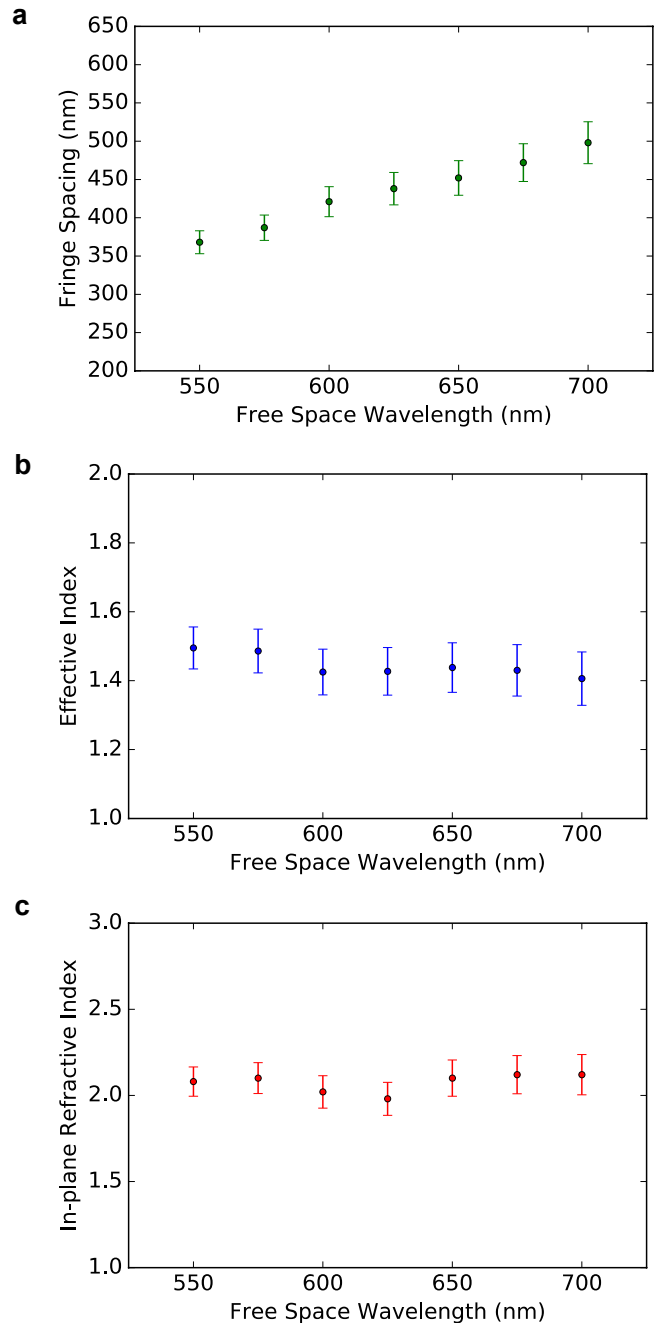

**Figure 3 | In-plane refractive index measurement.** (**a**) Observed interference fringe spacing as a function of the wavelength. (**b**) Corresponding effective index of the waveguide mode. (**c**) Extracted values for the in-plane refractive index of h-BN as a function of wavelength.

The spatial periodicity of the fringes can be found using a 2D FFT directly on the NSOM images (insets in Figure 2a-c), providing a direct measure of the guided wavelength (Figure 3a) and of the normalized effective index $n_{eff,\text{TE}}$ of the waveguide mode (Figure 3b) as a function of the free space wavelength. Due to TE nature of these modes, their dispersion is independent of the value of $n_z$, and it depends only on the in-plane refractive index, $n_x$, and on the thickness of the sample. A series of numerical simulations with different $n_x$ values can be used to obtain the relation $n_{eff,\text{TE}}(n_x)$, given by $\lambda = \lambda_0/n_{eff}$. Figure 3c shows the extracted $n_x$ of h-BN in the range 550-750 nm. The in-plane refractive index over the wavelength range 550-700 nm is in the range 1.98-2.12, corroborating previous studies of the refractive index of h-BN in the visible portion of the spectrum [24]. Observation of interference fringes in both the simulations and NSOM measurements indicates the interference of waveguide modes with the impinging x-axis polarized beam.

The second experiment relies on the use of SPPs, which are TM electromagnetic modes confined at the interface of a metal and a dielectric. They are characterized by an elliptically polarized evanescent electric field and by a well-known dispersion relation dictated by the complex relative electrical permittivities of the dielectric and the metal:

$$k_{SPP} = k_0 \sqrt{\frac{\epsilon_m \epsilon_d}{\epsilon_m + \epsilon_d}} \quad (1)$$

where $k_{SPP}$ is the wavevector of the surface plasmon, $k_0$ is the free space wavevector, $\epsilon_m$ is the complex relative electrical permittivity of the metal, and $\epsilon_d$ is the complex relative electrical permittivity of the dielectric. However, this dispersion relation only holds when the dielectric is isotropic and semi-infinite. For the case of a thin anisotropic dielectric such as h-BN, the dispersion relationship is affected by the anisotropy, as the polarization has both in-plane and out-of-plane components (see Figure 1d). Thus, propagation depends on both the in-plane and out-of-plane dielectric constants and can be used to probe them. The dispersion relation of SPPs in the presence of an anisotropic and finite thickness dielectric can be derived as [35]:

$$\left(\frac{k_a}{\epsilon_x}+\frac{k_d}{\epsilon_d}\right)\left(\frac{k_a}{\epsilon_x}+\frac{k_m}{\epsilon_m}\right)=\left(\frac{k_a}{\epsilon_x}-\frac{k_d}{\epsilon_d}\right)\left(\frac{k_a}{\epsilon_x}-\frac{k_m}{\epsilon_m}\right)\cdot\exp(-2k_a d) \quad (2)$$

with:

$$k_a=\sqrt{\epsilon_x\left(\frac{k_x^2}{\epsilon_z}-\frac{\omega^2}{c^2}\right)}, \quad k_d=\sqrt{k_x^2-\epsilon_d\frac{\omega^2}{c^2}}, \quad k_m=\sqrt{k_x^2-\epsilon_m\frac{\omega^2}{c^2}} \quad (3)$$

where, $\epsilon_x$ is the complex relative electrical permittivity of h-BN along $x$, $\epsilon_z$ is the complex relative electrical permittivity of h-BN along $z$, $\epsilon_d$ is the complex relative electrical permittivity of the dielectric superstrate above h-BN which in our case is air (i.e., $\epsilon_d = 1$), $\epsilon_m$ is the complex relative electrical permittivity of metal, $k_x$ is the in-plane SPP wavevector, $\omega$ is the angular frequency, $c$ is the speed of light, $d$ is the thickness of h-BN, $\epsilon_x = n_x^2$ and $\epsilon_z = n_z^2$. Essentially, a measurement of the effective mode index of the elliptically polarized SPPs in the presence of 10 nm thin anisotropic h-BN, will allow for the extraction of the out-of-plane refractive index, given that the in-plane refractive index is known a priori from the waveguide modes.

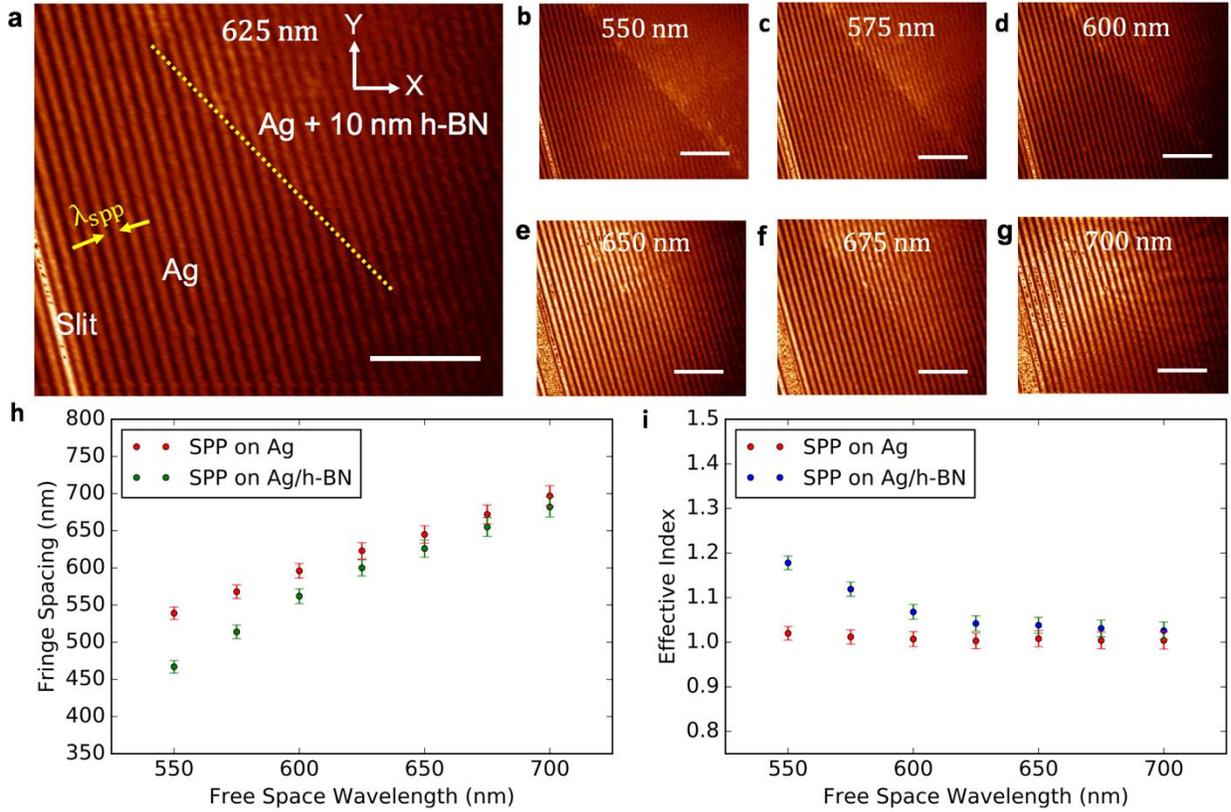

**Figure 4 | NSOM observation of TM SPP modes in h-BN on Ag**. (**a-g**) NSOM scans for $\lambda_0$ in the spectral range 550-700 nm. The slit is positioned in the bottom left corner (not fully pictured) where the optical signal on the detector saturates. The 10 $nm$ h-BN flake is visible from the middle to the right side of the image, as depicted in (**a**) by a yellow dashed line. (**h**) Observed SPP interference fringe spacing inside and outside the region covered by the flake. (**i**) Corresponding SPP effective index, given by $n_{eff} = \lambda_0/\lambda_{SPP}$. Scale bars in (**a-g**) are 5 μm.

Figure 4 summarizes the NSOM experimental results for the near-field SPP experiment. The FIB milled slit is illuminated from the back-side with light polarized perpendicularly to it, launching SPPs with wavefronts parallel to the slit. The initial propagation of the SPPs on the Ag is altered by the presence of the thin film h-BN, which changes the SPP wavelength, $\lambda_{SPP}$, due to the different refractive indices. The interference fringes (Figure 4a-g) are spaced by the SPP wavelength, $\lambda_{SPP}$. Following the same argument as in the first experiment, an FFT can be used to obtain $\lambda_{SPP}$ and the corresponding effective index $n_{eff,SPP}$ (Figure 4h,i). A calibration factor of 1.03 is applied to the data collected at the Ag/h-BN/air interface in an effort to account for the systematic errors before the experimental data are compared to the simulation results (see Supplementary Information for detailed description). Strikingly, despite the very low number of h-BN layers, the effect on the SPP propagation is evident, and allows for the extraction of useful information on the sample.

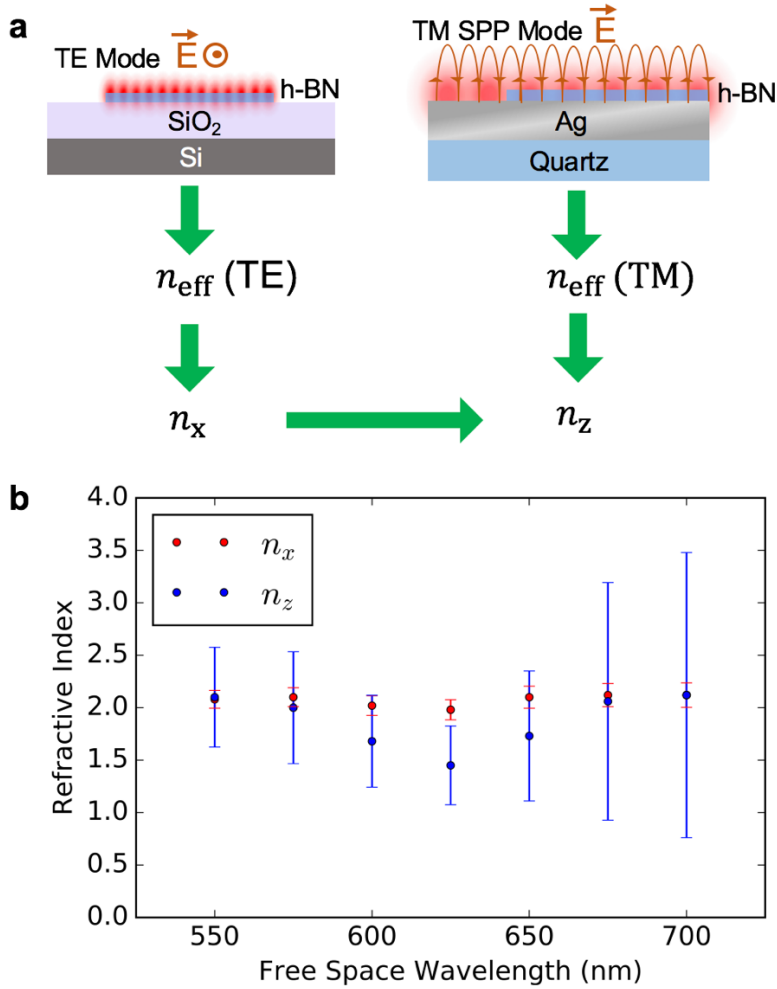

**Figure 5 | Extraction of anisotropic refractive indices of h-BN**. (**a**) Procedure to extract the anisotropic refractive indices: the in-plane component $n_x$ is first extracted by matching the experimental effective index $n_{eff}$ of the TE mode with the one obtained from numerical FDTD simulation. Thereafter, the out-of-plane $n_z$ component is found from the numerical FDTD simulation by knowing: the thickness from atomic force microscopy measurements and the value of $n_x$ previously extracted from the waveguide mode case and the experimental $n_{eff}$ of the SPP in the presence of h-BN. This procedure is repeated for all wavelengths. (**b**) Anisotropic refractive indices of h-BN in the spectral range 550-700 nm. Error bars represent the uncertainty due to the limited number of fringes accessible in the NSOM images (see Supplementary Information for detailed description of error bar calculation).

Because the SPP propagation in presence of the h-BN is affected by both $n_x$ and $n_z$, this experiment alone is insufficient to extract the value of $n_z$. However, using the previously measured $n_x$ in the first experiment, $n_z$ can be extracted from FDTD simulations, as outlined in Figure 5a. Figure 5b shows the resulting in-plane and out-of-plane refractive index values of h-BN. Over the wavelength range 550-700 nm, $n_z$ is in the range 1.45-2.12 and $n_x$ is in the range 1.98-2.12.

**Conclusions**

2D materials are expected to exhibit new physics when approaching the monolayer limit [9, 13, 17-20]. However, the ability to experimentally measure the anisotropic refractive indices of a thin, small area material is traditionally a difficult experiment, even for isotropic materials. Our methodology allows us to measure two unknowns—the in-plane and out-of-plane refractive indices of h-BN, which has been used in several works as a dielectric material [13, 21, 22]. The strategic material choice of h-BN for the proof of principle has two main benefits: previous experiments have shown thin film h-BN to not differ from bulk h-BN too drastically, so the data can be checked against previous experiments and, most importantly, it paves the way for the measurement of materials that require h-BN encapsulation, such as transition metal dichalcogenides [9, 18-20] and black phosphorous [8, 9, 15]. We anticipate that the knowledge of the optical constants of 2D materials, and particularly the possibility of experimentally accessing their out-of-plane optical properties due to the SPP coupling, will allow for their use in a wider variety of optics experiments and devices.

**Materials and Methods**

**Fabrication:**

*Devices on SiO$_2$:* h-BN is mechanically exfoliated on to a 285nm SiO$_2$/Si substrate with pre-defined metallic alignment marks. The substrates are then coated with MA-N 2403 (negative electron beam resist) and exposed with an electron beam system with a dose of 1200 μC/cm$^2$ using an accelerating voltage of 125 kV. The samples are shaped into rectangles, triangles, or circles. After developing in AZ-726 for 1 minute, the samples are post-baked at 100 C for 10 minutes. Then h-BN is etched by using a reactive ion etching (RIE) system with CHF$_3$/Ar/O$_2$ at flows of 10/5/2 sccm respectively and a RF generator at 30 W for 2-5 minutes. After the etching process, exposed MA-N 2403 resist is removed by Remover PG and chloroform. Afterwards, the samples are rinsed with isopropyl alcohol (IPA) and dried with nitrogen.

*Devices on Ag*: A single-crystal 10 nm thin h-BN flake is aligned and transferred to the active device region on the Ag substrate using a standard dry transfer technique [4].

**Numerical Simulations:**

*Full wave 3D simulations:* The Lumerical FDTD 3D wave solver is used to compute the fields in the patterned shapes considered in the experiment. For the case of waveguide modes in 65 nm thick h-BN on top of 285 nm SiO$_2$/Si, a fine mesh size of 1 nm is used. The monitor is placed 10 nm above the h-BN layer to compute the near-field power of the waveguide mode. A Gaussian source of 650 nm is placed above the geometry.

*Effective indices simulation and extraction:* The 2D solver Lumerical Mode Solutions is used to compute the mode profile and effective indices for a range of refractive indices of h-BN. For the case of waveguide mode where 65 nm thin h-BN is placed on top of 285 nm SiO$_2$/Si, h-BN is modeled as an isotropic dielectric, since the propagation is independent of the out-of-plane properties of h-BN. A range of refractive indices for h-BN is swept across to obtain the corresponding effective indices. The refractive index for which the corresponding computed effective index matches that of the experimental effective index measured from NSOM scans, is assigned as the in-plane refractive index of h-BN for that particular wavelength. This process is repeated for other wavelengths. On the other hand, for the case of SPP mode where 10 nm thin h-BN is placed on 50 nm Ag substrate, h-BN is modeled as an anisotropic dielectric and in-plane refractive index from the waveguide mode is used as an input. Thereafter, the out-of-plane refractive index of h-BN is swept across a range to obtain the corresponding effective indices. The out-of-plane refractive index for which the corresponding computed effective index matches that of the experimental effective index measured from NSOM scans, is assigned as the out-of-plane refractive index of h-BN for that particular wavelength. Similarly, this process is repeated for other wavelengths.

## Acknowledgments

This work was supported by the NSF EFRI, award no. 1542807. This work was performed in part at the Harvard University Center for Nanoscale Systems (CNS), a member of the National Nanotechnology Coordinated Infrastructure Network (NNCI), which is supported by the National Science Foundation under NSF ECCS award no. 1541959. M.T. acknowledges the support of the Swiss National Science Foundation (SNSF) grant no. 168545. K.W. and T.T. acknowledge support from the Elemental Strategy Initiative conducted by the MEXT, Japan and JSPS KAKENHI grant no. JP15K21722.

## Author contributions

K.W. and T.T provided the h-BN crystals. K.W., L.J., K.P., and K.C., exfoliated the flakes, transferred it to the substrate and performed the lithography. K.C., A.Y.Z., D.W., R.C.D., and J.C. contributed to the fabrication of the plasmonic substrate. K.C., D.W., A.Y.Z., and A.A. performed the NSOM measurements. K.C., D.W., A.A., and M.T. analyzed the data. D.W., K.C., M.T., P.K., and F.C. wrote the paper with extensive inputs from all authors. D.W., P.K., and F.C. led the project.

# Supplementary Information for "Guided Modes of Anisotropic van der Waals Materials Investigated by Near-Field Scanning Optical Microscopy"

Daniel Wintz[1‡], Kundan Chaudhary[1‡], Ke Wang[2], Luis A. Jauregui[2], Antonio Ambrosio[3], Michele Tamagnone[1], Alexander Y. Zhu[1], Robert C. Devlin[1], Jesse D. Crossno[1], Kateryna Pistunova[2], Kenji Watanabe[4], Takashi Taniguchi[4], Philip Kim[2], Federico Capasso[1]

[1]Harvard John A. Paulson School of Engineering and Applied Sciences, Harvard University, Cambridge, MA 02138, USA

[2]Department of Physics, Harvard University, Cambridge, MA 02138, USA

[3]Center for Nanoscale Systems, Harvard University, Cambridge, MA 02138, USA

[4]National Institute for Materials Science, Namiki 1-1, Tsukuba, Ibaraki 305-0044, Japan

‡ These authors contributed equally to this work

Corresponding author: capasso@seas.harvard.edu

## Supplementary Methods

**Mathematical Description of the Interference Fringes in Waveguiding Structures**

The origin of the interference fringes can be understood from a simple mathematical argument. The total electric field $\boldsymbol{E_{tot}}$ on the tip as it scans the sample is the sum of a direct contribution of the incident laser beam (either reflected from the first sample or transmitted through the second one) and the guided modes in h-BN. Because the Gaussian beam is focused on the sample, its phase on the sample surface is constant, and its electric field can be written as $\boldsymbol{E_0}$ (here, we will neglect the Airy discs which are simply intensity modulations without phase variations, and do not affect the following argument). The phase of the electric field of the guided modes, however, depends on the position, according to the direction of the in-plane mode. Therefore the electric fields can be written as $\boldsymbol{E_n} e^{i\boldsymbol{k_n} \cdot \boldsymbol{r}}$ where the subscript **n** refers to each guided mode, $\boldsymbol{k_n}$ is the in plane guided wavenumber and $\boldsymbol{r}$ is the position. For the SPPs sample there is a single guided mode launched by the slit, while for the patterned samples on SiO$_2$ each edge can launch a separate mode. Each guided mode has the same effective index $n_{eff}$ and therefore we have $|\boldsymbol{k_n}| = \frac{2\pi n_{eff}}{\lambda}$.

The collected intensity can be expressed as the squared absolute value of $\boldsymbol{E_{tot}}$:

$$|\boldsymbol{E_{tot}}|^2 = |\sum_{n=0}^{N} \boldsymbol{E_n}|^2 = |\boldsymbol{E_0} + \sum_{n=1}^{N} \boldsymbol{E_n} e^{i\boldsymbol{k_n} \cdot \boldsymbol{r}}|^2 = A + B + C \quad \text{(S1)}$$

$$A \equiv \sum_{n=0}^{N} |\boldsymbol{E_n}|^2 \quad \text{(S2)}$$

$$B \equiv 2\text{Re}\left(\sum_{n=1}^{N} \boldsymbol{E_n} \boldsymbol{E_0^*} e^{i\boldsymbol{k_n} \cdot \boldsymbol{r}}\right) \quad \text{(S3)}$$

$$C \equiv \text{Re}\left(\sum_{n=1}^{N} \sum_{m=1, m \neq n}^{N} \boldsymbol{E_n} \boldsymbol{E_m^*} e^{i(\boldsymbol{k_n} - \boldsymbol{k_m}) \cdot \boldsymbol{r}}\right) \quad \text{(S4)}$$

where $N$ is the number of guided modes in the sample. The summation $A$ gives a constant background in the plane. The summation $B$ represents the direct interference between each guided mode and the reflected beam, and which can be written as:

$$\sum_{n=1}^{N} 2|E_n||E_0| \cos(\theta_n - \theta_0) \cos(\varphi_n - \varphi_0 + \boldsymbol{k_n} \cdot \boldsymbol{r}) \quad \text{(S5)}$$

where $\theta_n$ represents the angle of the linear polarization and $E_n$ the complex phasor of the electric field. $\varphi_n$ represents the phase of the complex phasor of the electric field $E_n$. The summation $C$ gives the interference among guided modes, and can be written as:

$$2\sum_{n=1}^{N} \sum_{m=1, m \neq n}^{N} \cos(\theta_n - \theta_m) \cos(\varphi_n - \varphi_m + (\boldsymbol{k_n} - \boldsymbol{k_m}) \cdot \boldsymbol{r}) \quad \text{(S6)}$$

The NSOM scans demonstrate fringes which are compatible with the second part of the summation, since they have a spatial periodicity of $\frac{\lambda}{n_{eff}}$. We can come to this conclusion by noting that the dominant fringes are parallel to the edges which launches them. In addition, any interpretation of the fringes as standing waves would be incorrect, since this would require a doubled spatial periodicity, while the observed periodicity is just 40% shorter than the free space wavelength (Figure 3b). Therefore, the observed fringes cannot be standing waves associated to the guided modes.

The dominance of these fringes is due to the fact that $E_0$ is significantly larger than $E_1$, $E_2$, ... However, the summation $C$ (direct interference among guided modes) can be observed as fainter dots in the 2D Fourier transform, especially in the inset of Figure 2b.

The presence of the factor $\cos(\theta_1 - \theta_0)$ indicates that the interference fringes are stronger when the mode is propagating orthogonally with respect to the incident beam polarization, so that the fields are parallel. In addition, each edge scatters the incident beam in the guided mode with different intensities depending on the angle between the edge and the incident polarization. Since only the component of the incident field parallel to the edge can couple to the guided mode, another $\cos(\theta_1 - \theta_0)$ factor appears, so that the fringes intensity is proportional to $\cos^2(\theta_1 - \theta_0)$.

**SPP Interference Fringes**

The fringes observed in the SPP experiment comes from the interference of incident beam transmitted through the sample and the propagating SPPs. If the wavefront of the incident beam is flat on the plane of the sample, then constructive interference will occur with the propagating SPPs and the bright fringes will be spaced by $\lambda_{SPP}$. If, however, the incident wavefront has some curvature, the fringes will be spaced by an amount greater than $\lambda_{SPP}$. To see why, let us consider the phase evolution of the incident beam and the SPPs. The SPPs accumulate phase as they propagate in the form: $\phi_{SPP} = 2\pi x/\lambda_{SPP}$, and the slope of the line is $k_{SPP}$ (we take $x = 0$ to be the position of the slit). If the impinging beam is incident at an angle, the phase accumulated on the plane of the sample will be given by $\phi = 2\pi x \sin\theta / \lambda_0 + \phi_0$, where $\theta$ is the angular deviation from a flat wavefront and $\phi_0$ is the phase relative to the SPPs at $x = 0$. This is the reason why the fringes are spaced by greater than $\lambda_{SPP}$ when the incident wavefront is not flat, as graphically shown in Supplementary Figure 6.

**Error Sources**

There are four non-obvious sources of error in our experiments. First, the excitation laser has a bandwidth of approximately $10\ nm$. Second, there is a limited sample area from the NSOM images which is used to calculate the error bars for the experimentally measured interference fringe spacing, effective mode indices, and extracted anisotropic refractive indices. Another source of error is the fact that the SPP propagation length is dispersive. For higher frequencies, closer to the plasma frequency, the incident electromagnetic field penetrates more into the metal leading to larger optical losses and results in shorter propagation lengths. The opposite is true for lower frequencies, which experience larger propagation lengths. Therefore, as the SPPs propagate, the shorter wavelengths in the bandwidth of the laser are selectively absorbed more than the larger wavelengths resulting in a larger $\lambda_{SPP}$. This artificially higher $\lambda_{SPP}$ causes a reduction in the measured effective mode index given by the relation: $n_{eff} = \lambda_0/\lambda_{SPP}$. The other source of error comes from the focusing of the incident beam onto the sample. If the beam is perfectly focused on the plane of the sample, the wavefront of the incident beam will be flat as it interferes with the guided modes. This scenario produces sinusoidal interference fringes with high intensity peaks where the guided modes and the flat incident wavefront are in phase, with the fringes separated by $\frac{\lambda_0}{n_{eff}}$. If, however, the incident beam is not perfectly focused on the plane of the sample, the resulting interference fringes will be spaced by a different amount due to the incident wavefront curvature that creates a local local tilt $\theta$.

In fact, the local tilt $\theta$ will introduce a variation in the fringe spacing that can be derived from the interference argument above considering an additional in-plane momentum $k_x = k_0 \sin(\theta)$. The spatial frequency of the fringes $k_{fringes}$ is then given by:

$$k_{fringes} = k_0 n_{eff} - k_0 \sin(\theta) = k_0 n_{eff}\left(1 - \frac{\sin(\theta)}{n_{eff}}\right) \tag{S7}$$

where the term $\frac{\sin(\theta)}{n_{eff}}$ is the relative error on the measured $n_{eff}$. Fortunately, this effect is systematic and can be corrected using two different approaches in the two experiments (waveguide modes and SPPs).

For waveguide modes, we perform the Fourier transform on fringes that are orthogonal to the Airy fringes of the Gaussian beam (Numerical aperture of the objective lens: 0.45, Rayleigh range: 210 nm), thus making sure that the error due to the local tilt is null.

Concerning the SPPs, since we have to focus on the slit to maximize the SPPs' intensity, the above strategy cannot be used because the fringes will always be parallel with respect to the Airy discs. Instead, the error can be compensated for by a calibration factor using the SPP fringe spacing on the Ag/air interface. The calibration factor is calculated by computing the ratio of the measured $\lambda_{SPP}$ to the value predicted by theory using data from McPeak 2015 [S1]. The average value for the calibration factor is 1.03 (all wavelengths have a calibration factor > 1), indicating a 3% increase in measured $\lambda_{SPP}$ as compared to the expected value from theory. This is consistent with the explanation that there are several sources of systematic error that cause an increase in the measured SPP wavelength.

**Error Bar Calculation:**

The error bars represent the uncertainty due to the limited number of fringes, and have been computed in the following way. First, we estimate the uncertainty of the effective indices of the TE waveguide modes ($n_{eff,TE}$) and of the SPPs ($n_{eff,SPP}$) from the limited resolution of FFT. Here, the relative uncertainty is given by:

$$\Delta n_{eff,TE} = \frac{n_{eff,TE}}{2N} \quad , \quad \Delta n_{eff,SPP} = \frac{n_{eff,SPP}}{2N} \tag{S8}$$

where N is the number of fringes in each case, which depends on the wavelength and on the size of useful portion of the NSOM image used in the FFT.

Subsequently, the uncertainty is propagated to find the errors on $n_x$ and $n_z$, namely $\Delta n_x$ and $\Delta n_z$. This is done using the standard propagation of errors in the non-linear case. In fact, noting that:

$$n_x = n_x(n_{eff,TE}, \lambda) \quad , \quad n_z = n_z(n_x, n_{eff,SPP}, \lambda) \tag{S9}$$

then:

$$\Delta n_x = \frac{\partial n_x}{\partial n_{eff,TE}} \Delta n_{eff,TE} \tag{S10}$$

$$\Delta n_z = \sqrt{\left(\frac{\partial n_z}{\partial n_x}\Delta n_x\right)^2 + \left(\frac{\partial n_z}{\partial n_{eff,SPP}}\Delta n_{eff,SPP}\right)^2} \tag{S11}$$

# Supplementary Figures

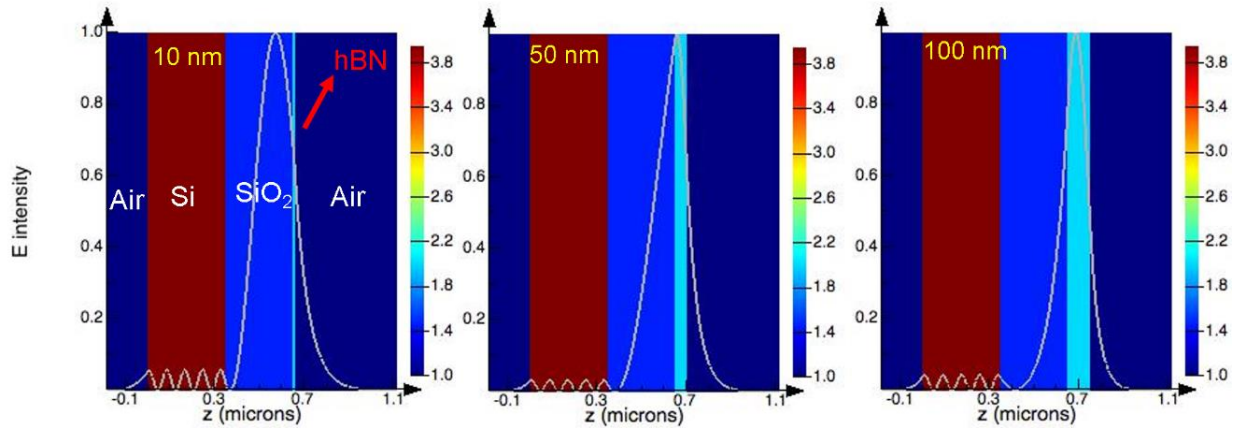

**Supplementary Figure 1 | 1D mode profile simulations for varying h-BN thicknesses**. $10 - 50 - 100\ nm$ h-BN thickness simulations at $\lambda_0 = 650\ nm$ (left to right), shown to visualize the modal volume present in the h-BN. The mode is TE.

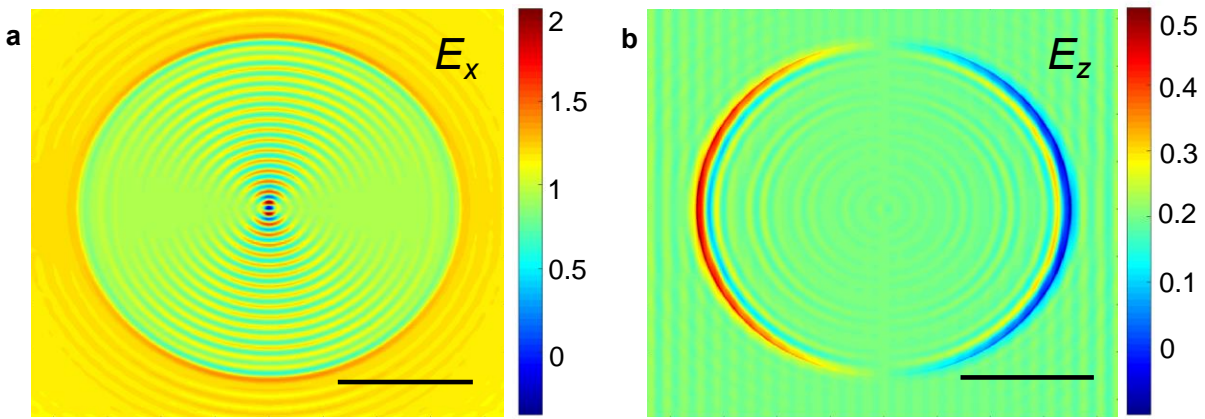

**Supplementary Figure 2 | FDTD simulations**. Full FDTD simulations at $\lambda_0 = 650\ nm$ highlighting the polarization of the h-BN guided modes, $E_x$ (**a**) and $E_z$ (**b**). The only portion of the electric field that is $z$-polarized is found at the edges where the incident beam is scattered in all directions. Scale bars in (**a**, **b**) are 5 μm.

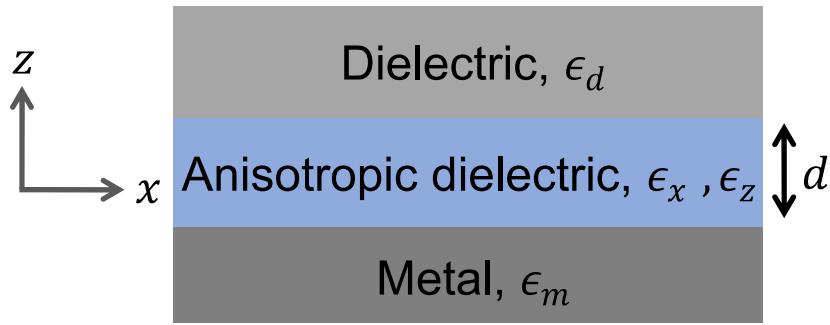

**Supplementary Figure 3 | Layered structure**. Full FDTD geometry used to derive the dispersion relationship.

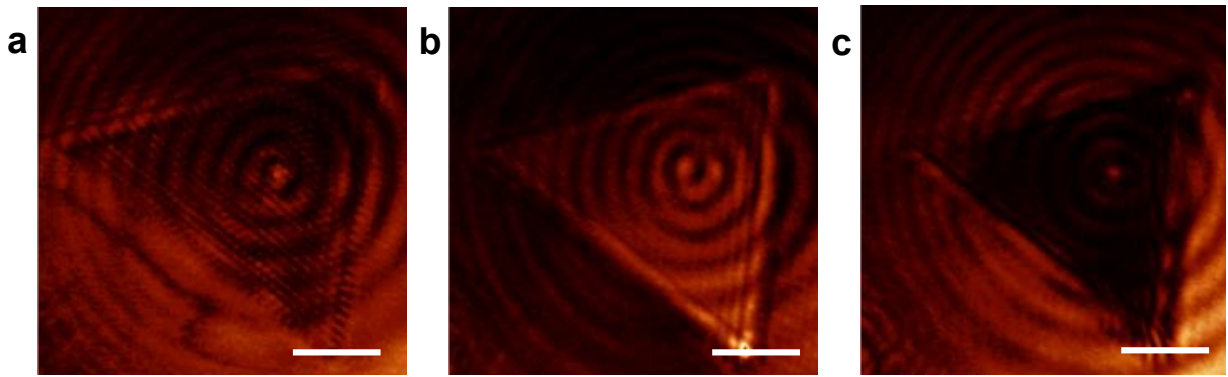

**Supplementary Figure 4 | Evanescent guided modes**. Triangular h-BN on SiO$_2$ is illuminated and scanned at various probe heights. **(a)** Roughly 10 $nm$, **(b)** 60 $nm$ and **(c)** 110 $nm$. Guided wave fringes are only clearly visible in **(a)** and completely dark in **(c).** Scale bars in **(a-c)** are 5 µm.

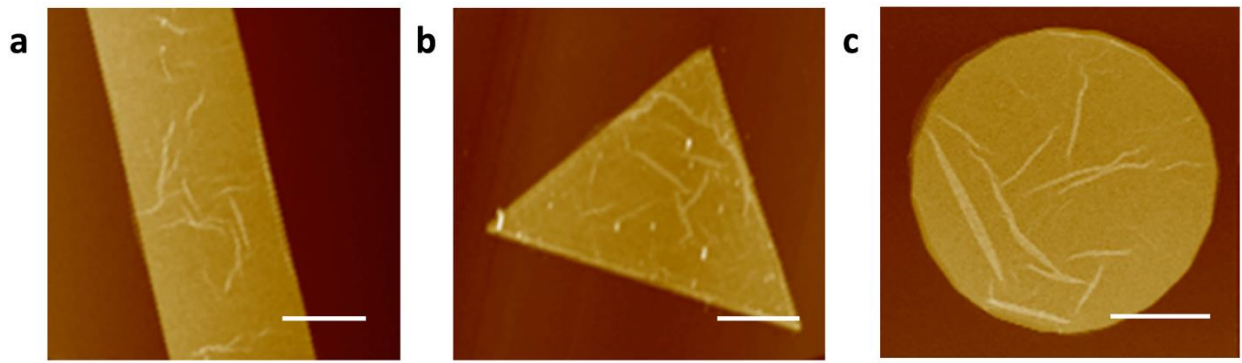

**Supplementary Figure 5 | Topography of various h-BN geometries. (a-c)** AFM topography of samples shown in **Figure 2a-c**. Scale bars in (**a-c**) are 5 μm.

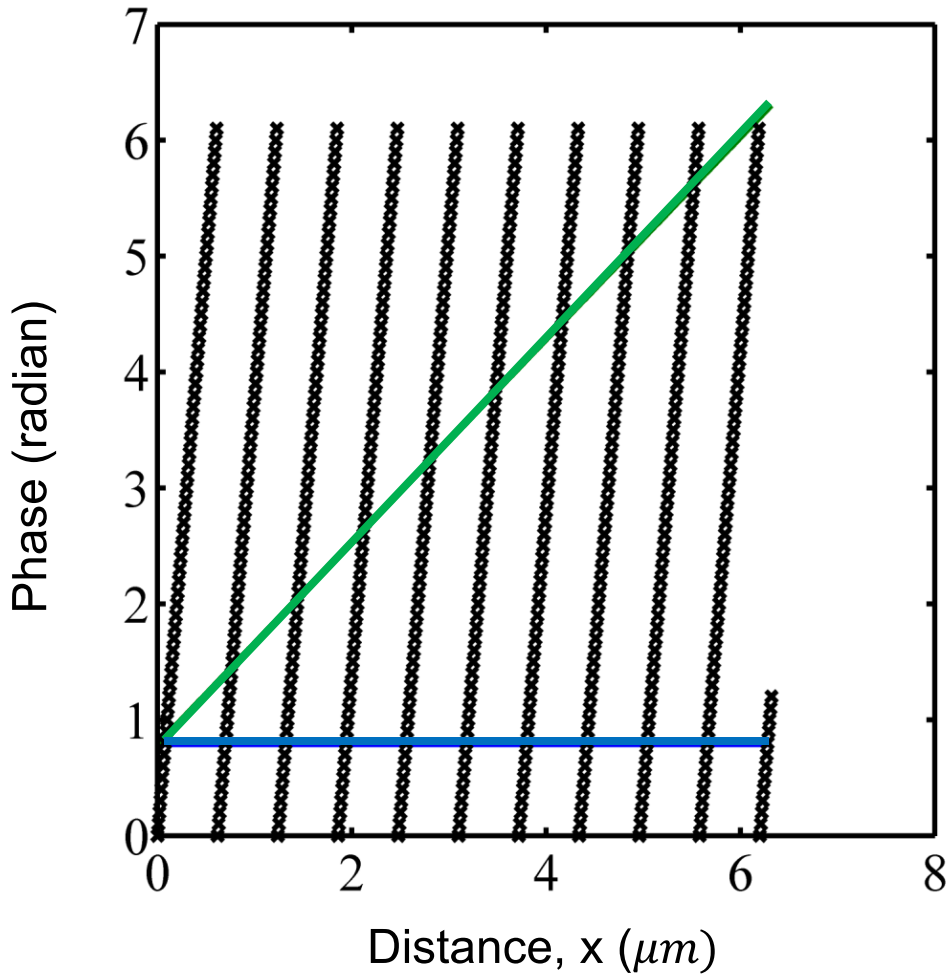

**Supplementary Figure 6 | Curved wavefront interference.** Black lines represent the phase accumulation (modulo $2\pi$) of the SPPs as they propagate away from the slit at $x = 0$. Blue line highlights the phase of a flat wavefront, and the intersections with the SPP phase will be the positions of constructive interference. The green line represents the phase accumulation of light incident on the sample at an angle of $\theta = 5°$, with $\phi_0 = \pi/4$. For $\lambda_0 = 632\ nm$, $\lambda_{SPP} = 620\ nm$. Geometric considerations prove that the interference fringe spacing for the light incident at an angle must be larger than $\lambda_{SPP}$.

# Supplementary References

Please refer to the main manuscript for references which do not begin with 'S'